\documentstyle[prb,twocolumn,aps,psfig]{revtex}
\tighten
\begin{document}
\draft
%\documentstyle[preprint,prb,aps]{revtex}
%\begin{document}
%\draft{}

\title{Cerenkov generation of high-frequency 
confined acoustic phonons in quantum wells}

\author{S. M. Komirenko, K. W. Kim}
\address{Department of Electrical and Computer Engineering, 
North Carolina State University, Raleigh, North 
Carolina 27695-7911}

\author{A. A.Demidenko, V. A. Kochelap} 
\address{Department of Theoretical Physics, Institute 
of Semiconductor Physics, National Academy of Sciences of Ukraine, Kiev-28, 
252650, Ukraine}

\author{M. A. Stroscio}
\address{U.S. Army Research Office, Research Triangular Park, NC 27709-2211}

\twocolumn[
%\date{\today}

\maketitle

\widetext \vspace*{-0.5 in}
\begin{abstract}
\begin{center} \parbox{14cm}{
We analyze the Cerenkov emission of high-frequency confined acoustic phonons by 
drifting electrons in a quantum well. We find that the electron drift can cause 
strong phonon amplification (generation). A general formula  for the gain 
coefficient $\alpha$ is obtained as a function of the phonon frequency and the 
structure parameters. The gain coefficient increases sharply in the short-wave 
region. For the example of a $Si/SiGe/Si$ device it is shown that the 
amplification coefficients of the order of hundreds of $cm^{-1}$ can be achieved 
in the sub-THz frequency range. 
}
\end{center}
\end{abstract}

]\narrowtext

High-frequency lattice vibrations with a high degree of spatial and temporal 
coherence have been observed for a number of semiconductor materials and 
heterostructures. These include Si, Ge, GaAs as well as SiGe and AlGaAs 
superlattices. \cite{r-6,merlin} These studies provide information on excitation 
mechanisms for the coherent phonons, their dynamics, electron-phonon interaction, 
and other important phenomena, including phonon control of the ionic motion.~\cite{Hu} 
Intense coherent phonon waves can be exploited for various applications: 
terahertz modulation of light, generation of high frequency electric oscillations, 
nondestructive testing of microstructures, etc. Usually, both optical and acoustic 
high-frequency coherent phonons are excited optically by ultrafast laser 
pulses.~\cite{r-6,merlin} The development of electrical methods of coherent 
phonon generation is an important problem.
\begin{figure}
\narrowtext
%\noindent
\psfig{figure=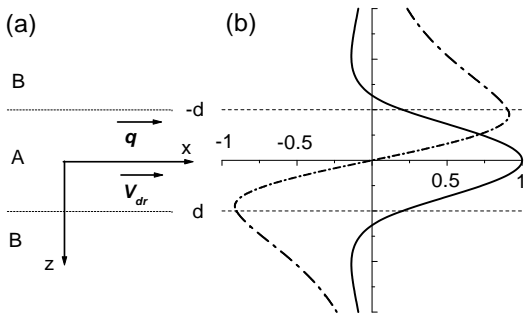,width=8.0cm,height=6.0cm,angle=270}
%{\epsfxsize=10.0cm\epsfysize=3.0cm \centerline{\epsfbox{Fig11.eps}}}
\protect \vspace{0.1cm}
\caption{(a) The analyzed heterostructure. (b) Distributions of the displacements 
$w_x$ and $w_z$ across the structure for the SSV mode calculated at $\alpha_{max}$ 
(see Fig. ~2.)}
\label{fig-1} \end{figure}
An electric current flowing though  a semiconductor can produce high-frequency coherent 
acoustic phonons. Two distinct cases are possible. If the current results from 
transitions of carriers between bound electron states, coherent phonon generation can 
occur if there is a population inversion between these states. Hopping vertical 
transport in superlattices and three barrier structures provides examples of mechanisms 
for the establishment of a population inversion and for stimulated generation of 
terahertz phonons~\cite{Tanya} and plasmons.~\cite{Gornik} If the current is due to 
free electron motion in an electric field, phonon amplification (generation) can be 
achieved via the Cherenkov effect if the electron drift velocity exceeds the 
velocity of sound. This effect is well-known for bulk samples.~\cite{sound}  High drift 
velocities and large densities of electrons are necessary for practical use of the 
Cerenkov effect. Advanced technology of semiconductor heterostructures opens new 
possibilities to employ this effect for high-frequency phonon generation. Indeed, such 
phenomena as high electron mobility at large electron density and phonon confinement 
in a quantum well (QW) can greatly facilitate achieving phonon amplification and 
generation by electron drift. In this letter, we analyze the generation of high-frequency 
confined acoustic phonons under the electron drift in a QW layer.

Consider a symmetric heterostructure shown in Fig.~\ref{fig-1}, (a) with electrons 
confined in the 
layer $A$ of thickness $2 d$. Assuming isotropic elastic properties for both semiconductors 
$A$ and $B$ one can introduce the longitudinal, $V_{LA}$ and $V_{LB}$, and transverse, 
$V_TA$ and $V_{TB}$, sound velocities. If $V_{TA} < V_{TB}, V_{LA},V_{LB}$, then localization 
of acoustic waves  near the embedded layer will occur.~\cite{Wendler,Mitin} 
These localized waves propagate along the layer and 
decay outside it. There are two classes of the localized waves: the shear-horizontal (SH) 
waves with the displacement vector $\vec u = (0,u_y,0)$, and the shear-vertical (SV) waves 
with the displacement vector: $\vec u = (u_x,0,u_z)$. Dispersion relations for each class 
of waves are represented by a set of branches $\omega=\omega_{\nu}(q)$, with $\omega$ and 
$q$ being the wave frequency and wave vector, $\nu$ is an integer. For a given $\nu$, 
localization of the waves depends on $q$. Let $\vec u_{\nu, q} (x,z,t) = \vec w_{\nu,q} 
(z) e^{(i q x - i \omega t)}$  be solutions of the elastic equations~\cite{Landau} 
describing the localized waves. Solutions with different "quantum  numbers" $\{\nu, q\}$ 
are orthogonal. We normalize the solutions by imposing the condition that the $\{\nu, q\}$ 
wave has an elastic energy equal to $\hbar \omega_{\nu,q}$.  The set of such solutions 
(modes) allows one to quantize the lattice vibrations, introduce {\em confined phonons} 
and analyze processes of absorption and emission of the phonons. 
 
Consider the interaction of a localized mode with electrons assuming that
a) only the lowest two-dimensional electron subband is populated and b) presence of 
higher subbands can be ignored. Then, setting the area of the layer equal to 1, the electron 
wavefunctions have the form $ \Psi_{\vec k} (\vec r, z) = e^{i \vec k \vec r}\chi (z)\,,$ 
where $\vec k$ is the two-dimensional electron wavevector. We suppose that electrons 
interact with phonons via the deformation potential (DP); thus, the energy of this 
interaction is $ H = b \,div\, \vec u $, where $b$ is the DP constant.  
Then the probability of transition between electron states $\vec  k$ and  $\vec k'$ due to 
emission or absorption  of a  confined phonon $\{\nu, q \}$  is 
$$
 P^{(\pm)} (k,k'|\nu, q) = \frac{2 \pi}{\hbar}
\left|M(q)\right|^2
\left( N_{\nu,q} + \frac{1}{2} \pm \frac{1}{2}\right)  \delta_{k_x \mp q,k'_x} $$
\begin{equation}\label{proba-1}
 \times \delta_{k_y,k'_y} 
\delta \left[E(\vec k) -E(\vec k') \mp \hbar \omega_{\nu,q}
\right]
 F(\vec k) \left[ 1 - F (\vec k' )\right],
\end{equation}
where $M (q)$ is the matrix element: 
\begin{equation} \label{matrix-1}
M (q) \equiv b~ \left( \int_{-\infty}^{\infty} div({\vec w}_{\nu,q }) 
\chi_1^2 (z) d z \right) / \kappa^{el} (q)\,,
\end{equation}
$N_{\nu,q}$ is the phonon number of the mode,
and $F(\vec k)= F[k_x,k_y]$ is the electron distribution function. In Eq.~(\ref{proba-1}) 
the upper signs correspond to emission, and the lower ones correspond  to  absorption 
processes. We take into account the effect of electron screening of the DP by introducing 
the electron permittivity:~\cite{Bastard}  $\kappa^{(el)} (q) = 1 + {2 \pi e^2 d} 
{\cal A} (q) {\cal B} (qd)/ {\kappa} $. Here, ${\cal A} (q)$ is the polarization operator 
of two-dimensional electrons: 
\begin{equation} \label{polarization}
{\cal A} (q) = - 2 \sum_{\vec k}
\frac{F (\vec k) - F(\vec k - \vec q)}{E(\vec k) - E(\vec k - \vec q)} ,
\end{equation}
where $\kappa$ is the dielectric constant and the factor ${\cal B} (s)$ is
$$ {\cal B} (s) = \frac{1}{s} \int_{-\infty}^{\infty}\int_{-\infty}^{\infty} d \zeta d \zeta' 
\chi^2 (\zeta d) \chi^2 (\zeta' d)
 e^{-s\, |\zeta -\zeta|}\,,
$$
Now, we introduce the kinetic equation for the phonon number of the mode $\{\nu,q\}$:
\begin{equation} \label{kinetic-1}
\frac{d N_{\nu,q}}{d t} = \gamma_{\nu,q}^{(+)} (1+N_{\nu,q})
-\gamma_{\nu,q}^{(-)} N_{\nu,q} - \beta_{\nu,q} N_{\nu,q}\,,
\end{equation}
where $\gamma_{\nu,q}^{(\pm)}$ are parameters which determine evolution of $N_{\nu, q}$ in 
time due to the interaction with electrons. Both parameters can 
be found easily by calculating the {\em total rates} of emission and absorption of  
phonons of a given mode by the summation of Eq.~(\ref{proba-1}) over all initial and final 
electron states. The parameter $\beta_{\nu,q}$ describes phonon losses. They can include  
phonon scattering or phonon absorption due to non-electronic mechanisms, phonon decay due 
to anharmonicity of the lattice, etc. In Eq.~(\ref{kinetic-1}) the terms which correspond to 
stimulated processes can be represented by $ \left(\gamma_{\nu,q}^{(+)} - \gamma_{\nu,q}^{(-)} 
\right) N_{\nu,q} \equiv \gamma_{\nu,q} N_{\nu,q}\,,$ with the phonon increment (decrement) 
equal to
\begin{equation} \label{increment}
\gamma_{\nu,q} =
\frac{m^*}{\pi \hbar^3 q} \left| M(q) \right|^2 
\left({\cal I}^{(+)} (q) - {\cal I}^{(-)} (q)\right)\,,
\end{equation}
\begin{equation} \label{I-q}
{\cal I}^{(\pm)} (q) = \int_{-  \infty}^{\infty} d k_y
F \left[ sign(q)\,\frac{m^* \, \omega_{\nu,q}}{\hbar |q|} \pm \frac{1}{2} q ,\, k_y
\right]\,.
\end{equation}
Here, $m^*$ is the effective mass. 

Depending on the shape of the electron distribution 
function, $F[k_x,k_y]$, the value $\gamma_{\nu,q}$ can be 
either positive, or negative.  If the phonon increment 
caused by the electron-phonon interaction  is positive 
and, in addition, it exceeds phonon losses, 
$\gamma_{\nu,q} > \beta_{\nu,q}$, the population of 
corresponding mode(s)  should increase in time,  
i.e., we obtain the effect of phonon generation.

One can introduce the amplification (absorption) coefficient for the confined acoustic  
modes which describes the rate of increase in the  acoustic wave  intensity per unit length. We 
obtain the amplification coefficient via the phonon increment: 
$\alpha_{\nu, q} = \gamma_{\nu,q} / V_g$, where   $V_g= d \omega_{\nu,q} / d q$ is the group 
velocity of the wave. The signs of $\gamma_{\nu,q}$ and $\alpha_{\nu, q} $ are determined by the 
factor $({\cal I}^{(+)} - {\cal I}^{(-)})$, which is to be calculated from the distribution 
function. This factor can be interpreted as the difference in the populations of the 
electron states, which are involved in the processes of emission and absorption. 
If this factor is positive, one obtains a kind of "population inversion".

We suppose that the electrons drift in an  applied electric field along the QW layer. Under the 
realistic  assumption of strong  electron-electron scattering, the distribution function can be  
thought as the shifted Fermi distribution:
\begin{equation} \label{distribution-1}
F[k_x,k_y] = F_F \left[ k_x - \frac{m^*}{\hbar} V_{dr}, \, k_y \right] \,,
\end{equation}
where $F_F (\vec k)$ is the  Fermi function, $V_{dr}$ and $T$ are the electron drift velocity 
and temperature, respectively. 
From Eq.~(\ref{increment}) for phonons propagating along the electron flux ($q >0$), we 
immediately find that $\gamma_{\nu, q},~\alpha_{\nu,q} >0$ if the electron drift  
velocity exceeds the confined phonon phase velocity:
$V_{dr} > \omega_{\nu,q}/| q|\,.$
This criterion is, in fact, the well-known condition of the Cerenkov generation effect.~\cite{sound} 
If $q < 0$ we always have  $\gamma_{\nu, q}\,,\,\, \alpha_{\nu,q} < 0$.

Typically, both velocities, $V_{dr}$ and $\omega_{\nu,q}/| q |$, are much less than the average 
electron  velocity. This implies that there is a relatively 
small disturbance of the Fermi function. Thus, to estimate  $\gamma_{\nu, q}$ and $\alpha_{\nu,q}$ we 
will take into account the shift in $F[k_x,k_y]$. While calculating the screening effect 
[see Eq.~(\ref{polarization})] we can neglect this shift and use just the Fermi function $F_F(\vec k 
)$. The latter approximation finalizes the description of amplification of the confined phonons by 
the drifting electrons. 

Now we shall apply these results to confined phonons of different symmetry. It is easy to see 
that the deformation-potential interaction couples only SV phonons with electrons. One can show 
that the functions $w_x (z)$ and $w_z (z)$ always have {\em different} symmetry. We define the 
symmetric shear-vertical (SSV) modes as those with $w_x(z)= w_x (-z)\,, w_z(z)=- w_z(-z)$ and  the 
antisymmetric ones with $w_x(z)= -w_x (-z)\,, w_z(z)= w_z(-z)$.  For a symmetric QW,  
the electrons are coupled with the SSV phonons. The displacement field distribution for 
one of the confined SSV modes is presented in Fig.~\ref{fig-1} (b).
\begin{figure}
\narrowtext
%\noindent
\psfig{figure=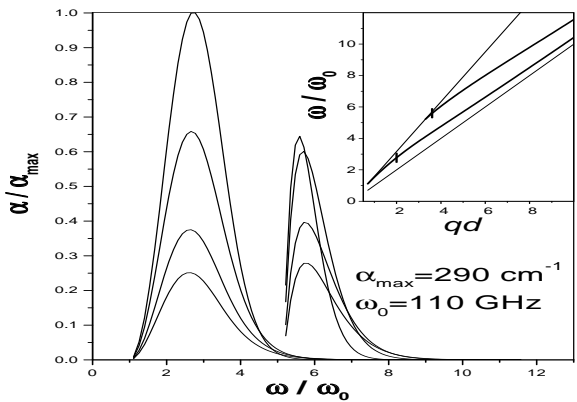,width=8.0cm,height=6.0cm}
%{\epsfxsize=10.0cm\epsfysize=3.0cm \centerline{\epsfbox{Fig11.eps}}}
%\protect \vspace{0.1cm}
\caption{The amplification coefficient versus  frequency for two lowest SSV phonon branches. 
T= 50, 100, 200 and 300 K. Increasing of T leads to decreasing of $\alpha$ at maximum. The 
values of $\omega$ and $q$ which correspond to the maxima of $\alpha$ at T=50 K are indicated on the 
dispersion relations shown in the Insert.}
\label{fig-2} \end{figure}
We have performed numerical calculations of the amplification coefficient for different heterostructures. 
We have found that two effects contribute critically to amplification: 
the phonon confinement effect through the 
matrix element of Eq.~(\ref{matrix-1}) and the 
nonequilibrium population of electron states through 
the factor $({\cal I}^{(+)} - {\cal I}^{(-)})$. For $III-V$ and SiGe heterostructures, the acoustic 
mismatch is typically small and the lowest mode is an antisymmetric SV mode. Consequently, 
all SSV modes have finite frequency onsets. This determines two important features:  
a low-frequency cut-off of the amplification and a nonmonotonous dependence of the matrix element 
$M(q(\omega))$. The population factor of Eq.~(\ref{I-q}) limits phonon amplification at high 
frequencies. As a result, amplification band for each SSV phonon branch is relatively narrow.  
Two typical amplification bands are illustrated in Fig.~\ref{fig-2}. These results are obtained for 
a  $p$-doped $Si/Si_{.5}Ge_{.5}/Si$ structure. The heavy hole subband is the lowest one in the strained $SiGe$ layer. 
We set $d = 5 \, nm$, the hole density is taken as $10^{12}cm^{-2}$ and the drift velocity is 
$V_{dr}=2.5\, V_{TA}$ with $V_{TA}=3.4\,10^5 cm/s$ for the $SiGe$ layer. One can see that 
amplification coefficient of the order of tens to hundreds $cm^{-1}$ can be achieved for 
confined modes in the sub-THz frequency range. 

These values of $\alpha$ are well above unavoidable phonon losses due to the effects of anharmonicity 
and scattering on isotopes. ~\cite{Tamura} The condition of phonon generation 
in a single passage device, $\alpha L_x \gg 1$, can be realized for reasonable extensions of the 
structure, $L_x$. At the maximum of amplification, the phonon wavelength equals $160\,\AA$ and the 
generated phonon flux is confined to a layer of thickness of about $200\,\AA$. Thus, a short-wavelength 
and highly-collimated  beam of the coherent phonons can be amplified and generated in perfect QW 
heterostructures.

In conclusion, we have found that the drift of two-dimensional electrons can result in Cerenkov 
instability of the phonon subsystem: the phonon modes confined near the QW layer and propagating 
along the electron flux are amplified.  The amplification coefficient for these modes has a sharp  
maximum in the sub-THz frequency range. The amplification coefficient 
can exceed   hundreds of $cm^{-1}$ for the mode almost confined within the QW layer. 
Our results suggest that a simple electrical method for generation of high-frequency coherent 
phonons can be developed on the basis of the Cerenkov effect. 	

This work was supported by the U.S. Army Research Office and the Ukrainian State
Foundation for 
Fundamental Researches.

\end{document}